\documentclass[aps,superscriptaddress,twocolumn,twoside,floatfix,pra,nofootinbib,a4paper]{revtex4-2}

\usepackage{times}
\usepackage{amsfonts}
\usepackage{amsmath}
\usepackage{amssymb}
\usepackage{amsthm}
\usepackage{multirow}
\usepackage[normalem]{ulem}
\usepackage[T1]{fontenc}
\usepackage{float}
\usepackage{mathtools}
\usepackage{bm}
\usepackage[UKenglish,cleanlook]{isodate}   
\usepackage{mathrsfs}
\usepackage{dsfont}

\usepackage{booktabs}

\usepackage{physics}

\usepackage{xcolor}
\definecolor{carmine}{RGB}{150,0,24}
\definecolor{mycolor1}{rgb}{1.00000,0.00000,1.00000}

\usepackage{natbib}
\usepackage[colorlinks=true,linkcolor=blue,citecolor=magenta,urlcolor=blue]{hyperref}

\usepackage{graphicx}
\usepackage{tikz-cd}

\begin{document}


\title{Binarisation of multi-outcome measurements in high-dimensional quantum correlation experiments}


\author{Armin Tavakoli}
\affiliation{Physics Department and NanoLund, Lund University, Box 118, 22100 Lund, Sweden.}
\author{Roope Uola}
\affiliation{Department of Physics and Astronomy, Uppsala University, Box 516, 751 20 Uppsala, Sweden}
\affiliation{Nordita, KTH Royal Institute of Technology and Stockholm University, Hannes Alfvéns väg 12, 10691 Stockholm, Sweden}
\affiliation{Department of Applied Physics, University of Geneva, 1211 Geneva, Switzerland}
\author{Jef Pauwels}
\affiliation{Department of Applied Physics, University of Geneva, 1211 Geneva, Switzerland}
\affiliation{Constructor Institute of Technology (CIT), Geneva, Switzerland}
\affiliation{Constructor University, 28759 Bremen, Germany}

\begin{abstract}
High-dimensional systems are an important frontier for photonic quantum correlation experiments. These correlation tests commonly prescribe measurements with many possible outcomes but they  are often implemented through  many individual binary-outcome measurements that use only a single-detector. Here, we discuss how this binarisation procedure for multi-outcome measurements can open a loophole, unless specific device-characterisation assumptions are satisfied. We highlight that correlation tests designed for multi-outcome measurements can be trivialised in binarised implementations and we then show how to accurately analyse binarised data to reveal its quantum features. For seminal types of correlation experiments, such as Bell inequality tests, steering tests and prepare-and-measure experiments, we find that binarisation may incur a sizable cost in the magnitude of quantum advantages. This emphasizes the importance of both accurate data analysis and implementing genuinely multi-outcome measurements in high-dimensional correlation experiments.
\end{abstract}


\maketitle


\section{Introduction}
Quantum systems featuring controlled superpositions of more than two classical states are known as high-dimensional quantum systems. They are an emerging resource for quantum information science in general and for quantum correlation experiments in particular. They enable both phenomena and resources that have no counterpart in qubit systems, for instance contextuality \cite{Budroni2022} and hyperentanglement \cite{Kwiat1997}. High dimension is also associated with stronger quantum features: it enhances significantly the noise robustness of entanglement and quantum steering \cite{Saunders2010, Bennet2012}. For quantum communication applications, it has the pivotal advantage of enhanced noise- and loss-tolerance for quantum key distribution \cite{Cerf2002}. Certification of high-dimensional quantum systems has gained much interest and such devices are increasingly relevant in photonic quantum technology; see e.g.~\cite{Dada2011, Gräfe2014, Malik2016, Kues2017, Erhard2018, Wang2018, Reimer2019, Bao2023}.

\begin{figure}[ht!]
    \centering
    \includegraphics[width=0.9\columnwidth]{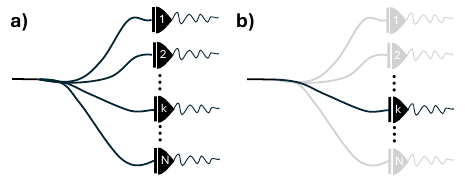}
    \caption{a) Multi-outcome measurement. All the $N$-possible outcomes are resolved. b) Binarised implementation of the multi-outcome measurement. A single detector is placed at port $k$. A click corresponds to the associated projector and a no-click to the complementary outcome. The detector is then placed at each of the $N$ ports one by one  in order to reconstruct the statistics of the multi-outcome measurement.}\label{Fig_binarised}
    \end{figure}

High-dimensional quantum protocols for seminal tasks such as entanglement detection, Bell nonlocality tests, Einstein-Podolsky-Rosen steering and quantum communication frequently prescribe measurements that can have many possible outcomes in every round of the experiment. However, in photonic platforms, these multi-outcome quantum measurements can be challenging to implement. It is therefore common practice to emulate multi-outcome measurements by employing a procedure in which  the quantum operation associated with each possible outcome is implemented separately. That is, instead of resolving many possible outcomes via equally many detectors, one makes a series of independent binary-outcome measurements, each corresponding to whether or not the system triggers a single detector. Thus, this binarisation procedure replaces a single high-dimensional $N$-outcome measurement with  $N$ separate projections corresponding to each of its possible outcomes. The multi-outcome measurement and the binarised implementation are illustrated in Fig.~\ref{Fig_binarised}.  

\begin{table*}[]
	\begin{tabular}{c|ccc}
		& \begin{tabular}[c]{@{}c@{}}Affected by binarised\\ measurements\end{tabular} & \begin{tabular}[c]{@{}c@{}}Quantum features in \\ binarised data\end{tabular} & \begin{tabular}[c]{@{}c@{}}Expected advantage from \\ multi-outcome measurements\end{tabular} \\ \hline
		Bell inequality test                                                             & Yes                                                                          & Yes                                                                           & Yes                                                                                          \\
		Steering test                                                                    & Yes                                                                          & Yes                                                                           & No                                                                                           \\
		\begin{tabular}[c]{@{}c@{}}Dimension witness \&\\ semi-DI protocols\end{tabular} & Yes                                                                          & Yes                                                                           & Yes                                                                                          \\
		Entanglement witness                                                             & No                                                                           & $-$                                                                             & $-$                                                                                            \\
		Standard QKD                                                                     & No                                                                           & $-$                                                                             & $-$                                                                                           
	\end{tabular}
	\caption{The impact of binarising multi-outcome measurements in selected types of correlation experiments. In cases where the binarisation loophole is opened, we show that the binarised data can still be used to certify  quantum features. For Bell inequality tests and dimension witness tests, we find that binarised measurements lead to a sizable cost in the strength of the signal, whereas for steering we oftentimes find that binarisation can be performed without a large loss of detection power.}\label{TabOverview}
\end{table*}

Here, we examine the consequences of binarising multi-outcome measurements in high-dimensional quantum correlation experiments. We begin by discussing its impact in correlation tests that are originally designed for multi-outcome measurements.  To this end, we formalise the  binarisation procedure and show that it can  faithfully substitute for multi-outcome measurements only when strict trust assumptions are imposed on the measurement devices. While in settings with trusted devices, e.g. entanglement detection, such assumptions are common, they are at odds with the basic requirements in for example Bell inequality tests, steering tests and the many established approaches to semi-device-independent quantum information processing which assume only partially characterized devices. We provide a simple overview of the types of quantum correlation experiments for which this binarisation loophole is relevant.  This leads us to the natural question of how to accurately analyse the data obtained from binarised implementations of multi-outcome measurements for demonstrating that an experiment defies classical limitations. We show how this can be achieved in a systematic way and apply it to well-known correlation tests based on entanglement and quantum communication respectively. Our results show that the targeted quantum features can survive the  binarisation of the multi-outcome measurement, but that can come at a significant cost in the noise-robustness of the signal. For Bell inequality experiments, we demonstrate this explicitly for the well-known CGLMP inequalities \cite{Collins2002} and similarly for the quantum communication task of random access coding \cite{Ambainis2002, Nayak1999}. In contrast, for steering experiments \cite{Uola2020} we find that binarisation can be significantly less costly. Our discussion shows how to systematically analyse data from binarised measurement implementations in photonic high-dimensional quantum correlation experiments and it demonstrates the inherent power of multi-outcome measurements in high-dimensional photonics.

\section{Binarisation of multi-outcome measurements}
Consider a $d$-dimensional quantum measurement described by a set of outcome operators $\mathbf{E}=\{E_k\}_{k\in [N]}$. For simplicity, we choose $\mathbf{E}$ to be a standard basis measurement, i.e.~$E_k=\ketbra{e_k}$ for $k=1,\ldots,d$ with $\braket{e_k}{e_{k'}}=\delta_{k,k'}$.  Let us denote by $p_\text{multi}$ the correlations obtained from measuring with $\mathbf{E}$  some state $\rho$. The statistics are described by Born's rule, $p_\text{multi}(k|\rho) = \tr(\rho E_k)$. A binarised implementation of $\mathbf{E}$ makes $d$ separate projections onto the basis vectors $\ket{e_k}$ for each outcome $k$. This corresponds to a set of binary-outcome measurements $\mathbf{F}=\{F_{1|k}, F_{\perp|k}\}_{k\in[d]}$, where  the ``1''-outcome corresponds to a successful event, i.e.~$F_{1|k}=E_k$.  It is clear that the original distribution $p_\text{multi}$, based on $\mathbf{E}$, can be simulated via the binarisation procedure; we have $p_\text{multi}(k|\rho)=\tr(\rho E_k) = \tr(\rho F_{1|k}) = p_\text{bin}(1|k,\rho)$.

One may naively suggest that this means that $\mathbf{E}$ can be faithfully substituted by $\mathbf{F}$ in any correlation experiment, but this is not in general accurate. The reason is that the substitution requires the assumption that the set of projectors $\{F_{1|k}\}_k$, which represent post-selecting on the successful outcome of $d$ different measurements,  perfectly corresponds to the projectors $\{E_k\}_k$ in the multi-outcome measurement. Thus, the set $\{F_{1|k}\}_k$ (which containts operators corresponding to different independent measurements) must by itself constitute a valid POVM, i.e.~ 
\begin{equation}\label{binaryassumption}
\sum_{k=1}^d F_{1|k}=\openone \,.
\end{equation}
This \textit{binarisation assumption} means that the quantum devices realising $\mathbf{F}$ must satisfy additional trust-relations. 

\begin{figure*}
	\centering
	\includegraphics[width=2\columnwidth]{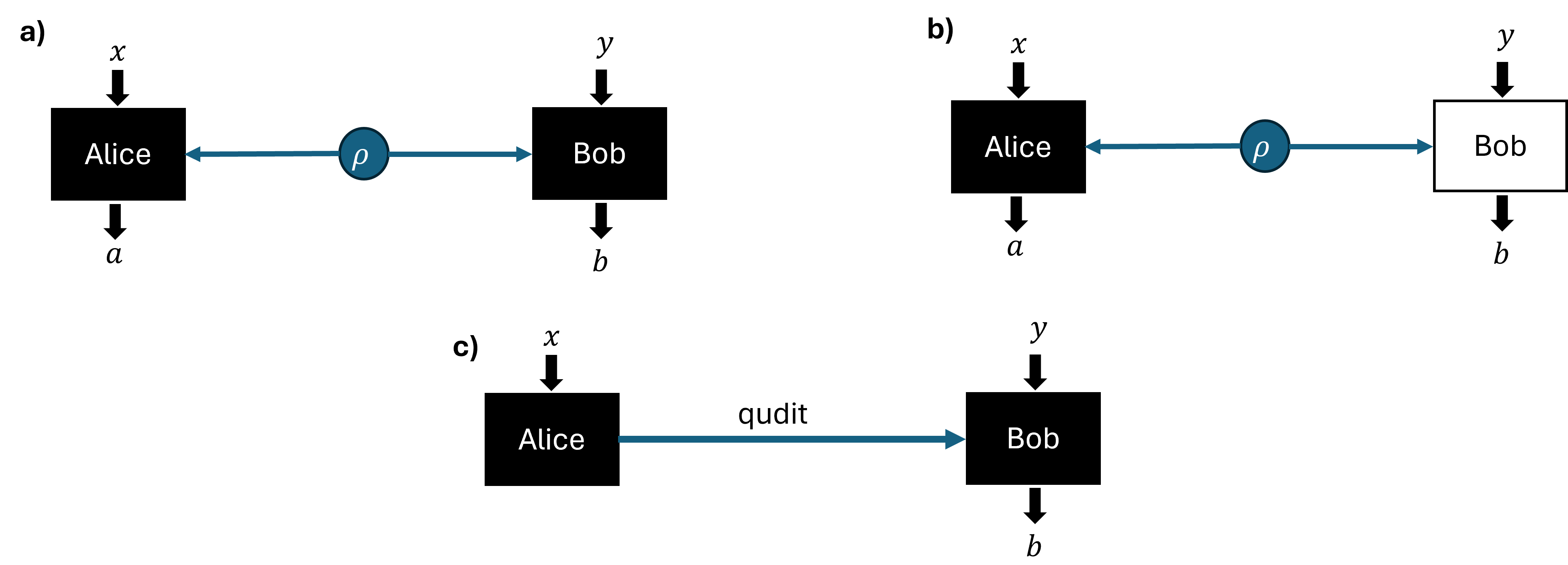}
	\caption{Selected correlation scenarios. a) Bell inequality scenario. b) Einstein-Podolsky-Rosen steering scenario. c) Dimension-witness scenario. Black boxes represent uncharacterised devices and white boxes represent fully characterised devices.}\label{FigScenarios}
\end{figure*}

Whether the binarisation assumption has an impact on the experiment depends on the nature of the theory under scrutiny by the experiment.  For instance, any experiment testing a device-independent property or protocol is at odds with the binarisation assumption. In a device-independent protocol, the multi-outcome measurement is considered unknown, i.e.~the measurement device is completely untrusted (it is a black box). The binarisation procedure therefore leads to an artificial situation in which the theoretical model stipulates that the measurement is limited only by quantum theory, whereas the experimental implementation implicitly assumes that the set of $d$ measurements used in the binarisation procedure satisfy the specific trust-relation \eqref{binaryassumption}. This lacks physical justification and overlooking it suggests the opening of a \textit{binarisation loophole} in the data analysis. Important types of entanglement-based correlation experiments that are concerned by this are Bell inequality tests \cite{Brunner2014, Tavakoli2022} and Einstein-Podolsky-Rosen steering \cite{Uola2020}, which are both based on black-box measurement devices; see Fig~\ref{FigScenarios}ab. Similarly, a variety of quantum communication schemes such as dense coding \cite{Bennett1992} and semi-device-independent quantum information processing schemes are concerned since these involve black-box measurements of prepared quantum states. For the same reasons, this applies also tests of the dimension of a physical system \cite{Hendrych2012, Ahrens2012, Martinez2018, Aguilar2018}, which are illustrated Fig~\ref{FigScenarios}c.

However, there are also types of correlation experiments that remain unaffected by using the binarisation assumption. These are scenarios in which the underlying theoretical model already assumes perfect control of the measurement device. In other words, the assumption \eqref{binaryassumption} is given. Important examples of this is high-dimensional generalisations of BB84's quantum key distribution protocol \cite{Sheridan2010} and entanglement witness tests \cite{Guhne2009}. Still, it is relevant that these types of tests are known to be vulnerable to implementational deviations from the perfect-trust assumptions \cite{Lydersen2010, Rosset2012} and taking these into account \cite{Morelli2022, Cao2024} would open up the binarisation loophole. In Table~\ref{TabOverview}, we  overview the impact of measurement binarisation for different types of quantum correlation experiments.

It is natural to examine the relationship between the binarisation loophole and the well-known detection loophole, which arises due to particle losses. When the setup is trusted, one can assume that successful detection events are representative also of the undetected events. This so-called fair sampling assumption serves as an analogue to the condition in Eq. \ref{binaryassumption}. However, when fair sampling cannot be assumed—such as in scenarios where the transmission line or measurement devices are not fully characterized—quantum features observed in the experiment could, in principle, be explained by classical models that selectively post-process successful detection events. This is known as the detection loophole \cite{Pearle1970,Clauser1974}.

In contrast, the binarisation loophole stems from the assumption that a collection of uncharacterized binary-outcome measurements faithfully represents a single multi-outcome measurement. Unlike the detection loophole, this issue persists even when the detection efficiency is 100\%, as it originates not from losses but from the measurement implementation itself. In practice, when  devices are uncharacterized, both the binarisation and detection loopholes must be considered independently.

As a simple example, consider the well-known $I_{3322}$ inequality, a Bell inequality where both parties measure a pair of three-outcome measurements. If no particles are lost in any part of the experimental setup, the detection loophole is closed. However, if one were to perform the three-outcome measurements through their binarisations, the relevant Bell inequality would be one for six binary measurements per side, i.e. an inequality for the $6622$ scenario. It is easy to convince oneself of the fact that if there is a local model for the $3322$ scenario, this model can be post-processed into a local model for the binarised $6622$ scenario. However, as we will show in the following sections, there are situations, where the binarised scenario has a local model, while the multi-outcome one violates a Bell inequality. Hence, even with the detection loophole closed, one can create a Bell inequality violation from fully local data set. This is the essence of the binarisation loophole.


\section{Quantum features from binarised measurements}
We now discuss how to accurately analyse the quantum properties of correlations obtained from binarised implementations of multi-outcome measurements (without relying on the binarisation assumption \eqref{binaryassumption}). Although we have seen that binarised measurements cannot be used in black-box tests designed for multi-outcome measurements, it may still be possible that their complete measurement statistics, $p_\text{bin}(1|k,\rho)$, defies classical models. In other words, there can exist classicality criteria tailored for binary-outcome measurements that are violated in the binarised experiment. To shed light on this, the key observation is that the correlations originally intended to be studied, based on the $N$-outcome measurement, should after binarisation instead be considered as arising in a different input-output scenario, in which outcomes are binary and the inputs are $N$ times more numerous. 

\begin{figure}[t!]
	\centering
	\includegraphics[width=0.95\columnwidth]{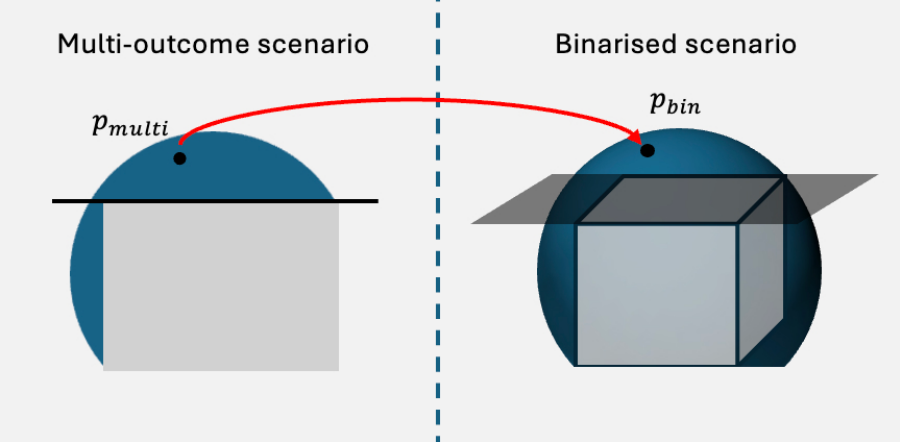}
	\caption{\label{fig:bin_corr} Accurate binarisation analysis. The distribution $p_\text{multi}$ belongs to the quantum set of correlations and can be detected as non-classical through a correlation inequality criterion (left). After binarising the distribution, we obtain $p_\text{bin}$, which belongs to a new correlation space associated with the revised input-output alphabets (right). To determine whether $p_\text{bin}$ is quantum, we must decide whether a classical model is possible in the new space, or alternatively determine a correlation inequality that detects its quantum features.}\label{Fig:sets}
\end{figure}

For example, consider a Bell experiment in which Alice and Bob each select from two ternary measurements on a shared state. Thus, in the original scenario they have two inputs and three outputs. However, if their measurement implementations are binarised, each ternary measurement is implemented as three binary measurements. Hence, their de-facto experiment corresponds to them having six inputs each and two outcomes each.  Indeed, since the correlations are established in a binarised experiment, one cannot apply correlation criteria designed for the multi-outcome scenario because the binarised correlations live in a different space. Fig.~\ref{Fig:sets} illustrates the need to view the correlations as arising in their proper input-output scenario. This leads to two natural questions. First, how do we in practice determine correlation criteria for binarised high-dimensional experiments? Secondly, will binarised quantum correlations still be sufficiently strong to demonstrate non-classical features? In what follows, we investigate these questions through illustrative case studies in nonlocality, prepare-and-measure and steering experiments. These three scenarios are illustrated in Fig~\ref{FigScenarios} and the results are summarised in Table~\ref{TabOverview}.

\subsection{Binarised nonlocality}
Consider a Bell experiment in which Alice's and Bob's inputs, $(x,y)$, have $M$ values and their outputs $(a,b)$ have $N$ values; see Figure~\ref{FigScenarios}a. The correlations in this scenario are denoted $p_\text{multi}(a,b|x,y)$. If both of them binarise their measurements, their respective inputs in the binarised experiment consist of tuples $\tilde{x}=(x,a)$ and $\tilde{y}=(y,b)$. Their joint distribution  becomes 
\begin{align}\label{bellcorr}\nonumber
	& p_\text{bin}(1,1|\tilde{x},\tilde{y})=p_\text{multi}(a,b|x,y) \,,\\\nonumber
	& p_\text{bin}(1,\perp|\tilde{x},\tilde{y})=p_\text{multi}^A(a|x)-p_\text{multi}(a,b|x,y) \,,\\\nonumber
	& p_\text{bin}(\perp,1|\tilde{x},\tilde{y})=p_\text{multi}^B(b|y)-p_\text{multi}(a,b|x,y) \,,\\\nonumber
	& p_\text{bin}(\perp,\perp|\tilde{x},\tilde{y})=1-p_\text{multi}^A(a|x)-p_\text{multi}^B(b|y)+p_\text{multi}(a,b|x,y) \, ,
\end{align}
where $p_\text{multi}^A$ and $p_\text{multi}^B$ are the marginal distributions of Alice and Bob, respectively.
The distribution $p_\text{bin}$ lives in a correlation space of a scenario in which both the parties have binary outputs and $M\times N$ inputs each. Thus, as in Fig.~\ref{Fig:sets}, we must transform from a $(M,N)$ input-output scenario into a $(M \times N,2)$ input-output scenario, in which we can study nonlocality.  Since the set of local correlations is a polytope \cite{Brunner2014}, the nonlocality of $p_\text{bin}$ can be decided using the standard linear programming characterisation of the local set \cite{Boyd2004}. The primal program can determine the distance between $p_\text{bin}$ and the local polytope, and the dual program provides a Bell inequality for witnessing the distribution \cite{SDPreview}.  

We carry out this analysis for the well-known Collins-Gisin-Linden-Massar-Popescu (CGLMP) Bell inequalities \cite{Collins2002}. These inequalities have binary inputs ($M=2$) and $d$ outputs per party. In the form of Ref.~\cite{Zohren2008}, they can be written
\begin{align}\nonumber
P(A_2&<B_2)+P(B_2<A_1)\\
&+P(A_1<B_1)+P(B_1\leq A_2)\geq 1 \, ,
\end{align}
where $A_j$ and $B_j$ are the outcomes of respectively Alice and Bob, for the $j$'th input.
We consider the optimal measurements \cite{Collins2002} applied to the optimal entangled state $\psi_N$ \cite{Zohren2008} mixed with white noise, i.e.~$\rho_v=v\psi_N+\frac{1-v}{N^2}\openone$ for visibility $v\in[0,1]$. From this, we get the optimal multi-outcome distribution $p_\text{multi}(v)$ mixed with white noise, from which we compute $p_\text{bin}(v)$. We then evaluate a linear program to determine the largest $v$ compatible with a local hidden variable model, for $N=2,\ldots,8$. Our implementation follows more efficient formulation outlined in Ref.~\cite{Bowles2021}. The results are summarized in Table~\ref{Tab_cglmp}. We see that $p_\text{bin}$ is nonlocal in all cases, and that the critical value of $v$ is significantly below unit, thereby permitting a significant signal. However, we also see that while the visibility decreases with $N$ in the original CGLMP Bell test, it instead increases in the binarised experiment. The results are similar if one instead of $\psi_N$ considers the maximally entangled state. This may be expected, since the cost of binarising measurements with increasingly many outcomes is higher. These results are in line with the intuition that binarisation gives additional power to the classical model under consideration; in this case local hidden variable theories.

\begin{table}[h!]
	\begin{tabular}{c|ccccccc}
		\hline
		CGLMP & \multicolumn{1}{c}{2} & \multicolumn{1}{c}{3} & \multicolumn{1}{c}{4} & \multicolumn{1}{c}{5} & \multicolumn{1}{c}{6} & \multicolumn{1}{c}{7} & 8    \\ \hline
		$v_\text{crit}^\text{multi}$ ($\%$)               & 70.7                                     & 68.6                                           & 67.3                   & 66.3                   & 65.6                   & 65.0                   & 64.5 \\ 
		$v_\text{crit}^\text{bin}$ ($\%$)                & 70.7                                      & 79.4                                           & 81.4                   & 83.4                   & 84.3                   &    85.3                    &     85.9  \\ 
	\end{tabular}
	\caption{Nonlocal correlations based on the CGLMP inequality using multi-outcome and binarised measurements. We consider the optimal quantum states and measurements for every $N=2,\ldots,8$ and evaluate the largest visibility of the state when mixed with white noise so that the correlations admit a local hidden variable model.}\label{Tab_cglmp}
\end{table}

\subsection{Binarised prepare-and-measure scenario}
Consider a prepare-and-measure scenario where Alice encodes input $x$ into a $d$-dimensional quantum state, $\rho_x$. It is sent to Bob who, depending on his input $y$, performs a measurement $\{E_{b|y}\}_y$ and records one of $N$ possible outcomes. This scenario is illustrated in Fig~\ref{FigScenarios}c. The correlations in the experiment are described by Born's rule, $p_\text{multi}(b|x,y)=\tr(\rho_xE_{b|y})$. However, if the measurement is implemented via binarisation, then we associate to Bob an input $\tilde{y}=(y,b)$ and a binary output $\{1,\perp\}$. The binarised distribution takes the form $p_\text{bin}(1|x,\tilde{y})=p_\text{multi}(b|x,y)$ and $p_\text{bin}(\perp|x,\tilde{y})=1-p_\text{multi}(b|x,y)$. To decide the classicality of $p_\text{bin}$ in the binarised input-output scenario, we can evaluate a suitable linear program \cite{Gallego2010, Renner2023}. To demonstrate this in practice, we have considered the task of high-dimensional random access coding since it has been the subject of several photonics experiments; see e.g.~\cite{Tavakoli2015, Aguilar2018, Farkas2021, Miao2022, Zhang2025}. 

In a random access code Alice's input consists of two $d$-valued symbols $x=x_1x_2\in\{1,\ldots,d\}^2$ and that Bob's marks which of them he wishes to learn by choosing  $y\in\{1,2\}$. Thus, the goal is for Bob to output $b=x_y$. The average success probability of the code is $\mathcal{P}_d=\frac{1}{2d^2}\sum_{x,y}p(b=x_y|x,y)$. For classical models, the success probability cannot exceed $\mathcal{P}_d\leq 1/2+1/(2d)$ but quantum models can increase this up to $\mathcal{P}_d\leq 1/2+1/(2\sqrt{d})$ \cite{Tavakoli2015, Farkas2019}. We focus on the simplest case of $d=3$ and binarise the probability distribution associated with the optimal quantum measurements and optimal quantum states when mixed with white noise. In the spirit of Fig.~\ref{Fig:sets}, this moves us from the input-output scenario associated with two measurements and three outcomes to a scenario associated with six measurements and two outcomes. From the linear program, we compute the maximal visibility compatible with a classical model and obtain $v_\text{crit}\approx 78.8\%$. We also obtain a correlation inequality tailor-made for being violated using binarised measurements on high-dimensional systems, 
\begin{equation}
	\sum_{x,y}\left( p_\text{bin}\left(1|x,(y,x_y)\right)-\frac{5}{8}\sum_{b\neq x_y} p_\text{bin}\left(1|x,(y,b)\right)\right) \leq 9.
\end{equation}
In order to assess the cost of binarisation, we can compare the visibility requirement to that obtained when using multi-outcome measurements. For the latter, we find $v_\text{crit}\approx 73.2\%$, which is significantly smaller. Thus, binarisation comes at a weakened signal, but the quantum features of the high-dimensional systems can still be detected.

\subsection{Binarised steering}
Consider now measurement binarisation in steering experiments. In the steering scenario (see Fig~\ref{FigScenarios}b) we need only consider the assemblage of states $\sigma_{a|x}^\text{multi}=\tr_A\left(A_{a|x}\otimes \openone \rho_{AB}\right)$ generated by Alice for Bob via her local measurements on the shared state. The binarised steering assemblage takes the form $\sigma_{1|\tilde{x}}^\text{bin}=\sigma_{a|x}^\text{multi}$ and $\sigma_{\perp|\tilde{x}}^\text{bin}=\rho_B-\sigma_{a|x}^\text{multi}$ where $\rho_B=\sum_{a}\sigma_{a|x}^\text{multi}$ is determined by the no-signaling condition. In analogy with Fig.~\ref{Fig:sets},  binarisation implies that the multi-outcome assemblage is mapped into a binarised assemblage with a larger input alphabet. To determine whether it is steerable, we must determine whether $\{\sigma_{\tilde{k}|\tilde{x}}^\text{bin}\}$ admits a local hidden state model in the binarised scenario. This can be achieved via a semidefinite program and the dual program provides an explicit steering inequality \cite{Cavalcanti2017}.

Recently, it has been shown that simple steering inequalities can be constructed for binarised steering assemblages based on $d$-dimensional mutually unbiased bases without any reduction in the critical visibility needed for a violation \cite{Srivastav2022}. This already suggests that binarisation in steering behaves different from the two previously considered scenarios. In fact, we find  that steering can often be binarised without any cost in visibility. Firstly, we have shown that for any pure state whose reduction has full rank, any steerable assemblage obtained from arbitrary projective measurements remains steerable after binarisation \cite{argument}. However, this is not expected to hold for general measurements and states because there exists pairs of incompatible qubit and qutrit measurements that have jointly measurable binarisations \cite{Reeb2013, JP2014, Uola2021subspace}.  Hence, one can produce steerable multi-outcome assemblages with unsteerable binarisations. Secondly, to further study binarised steering, we have conducted numerical tests based on random two-qutrit states and random projective measurements. These tests many times show that binarisation of steering can be performed without any cost in the noise tolerance of the entangled state, and at other times only with a small reduction in the noise tolerance. Note that for the set of all binary measurements performed on some important classes od states, a gap exists between binary and generic multi-outcome measurements ~\cite{Chaubin}.

\section{Conclusion}
Multi-outcome photonic measurements are often implemented by independently performing several single-detector measurements, each projecting onto one outcome or its complement. We have referred to such implementations as binarised measurement procedures and examined their impact on experiments probing quantum correlations. Our discussion highlights that binarised measurements introduce a loophole in correlation experiments designed to test theories involving uncharacterized multi-outcome measurements. Such measurements are fundamental to high-dimensional quantum systems, which are becoming increasingly significant in quantum information science. Leaving the binarisation loophole open can lead to false positives, incorrectly attributing quantum features to an experiment.

We have demonstrated two distinct approaches to closing this loophole. First, an accurate data analysis can be performed for binarised experiments. We have systematically shown how this can be achieved for key experimental paradigms such as Bell tests, steering tests, and quantum communication. This approach yields correlation criteria specifically tailored for binarised experiments. However, while quantum features often persist under binarisation, the degree of non-classicality they exhibit is generally reduced. This trade-off counterbalances the noise and loss resilience that often motivates the use of high-dimensional quantum systems in the first place.

Alternatively, our analysis underscores the importance of developing and implementing genuine multi-outcome measurements for high-dimensional quantum systems. Such advancements would unlock the full potential of high-dimensional quantum information, rendering the drawbacks of binarised measurement implementations irrelevant.

\begin{acknowledgments}
We are grateful for fruitful discussions with Otfried Gühne and Juha-Pekka Pellonpää regarding quantum measurement theory and steering using binary measurements. This work is supported by the Wenner-Gren Foundations, by the Knut and Alice Wallenberg Foundation through the Wallenberg Center for Quantum Technology (WACQT), the Swedish Research Council under Contract No.~2023-03498, the Wallenberg Initiative on Networks and Quantum Information (WINQ), and the Swiss National Science Foundation (Ambizione PZ00P2-202179) and NCCR-SwissMap.
\end{acknowledgments}

\bibliography{references_binarisation}

\end{document}